# API Phonons: Python Interfaces for Phonon Transport Modeling

Xin Qian[1*], Guanda Quan[1], Te-Huan Liu[1*], and Ronggui Yang[1,2*]

1 School of Energy and Power Engineering, Huazhong University of Science and Technology, Wuhan 430074, China

2 College of Engineering, Peking University, Beijing 100871, China.

* Corresponding Emails: xinqian21@hust.edu.cn, thliu@hust.edu.cn, ronggui@pku.edu.cn

**Abstract**

API Phonons is a Python software package to predict the transport dynamics of heat-carrying phonons. Using the powerful syntax of Python, this package provides modules and functions interfacing between different packages for atomistic simulations, lattice dynamics, and phonon-phonon interaction calculations including LAMMPS, Quippy, Phonopy, and ShengBTE. API Phonons enabled complex phonon calculations, including (1) extracting harmonic and anharmonic force constants from arbitrary interatomic potentials, which can be used as inputs for solving Boltzmann transport equations; (2) predicting thermal conductivity using Kubo's linear response theory, which captures both quasiparticle transport and inter-band coherent transport; and (3) modeling of ultrafast pump-probe thermal responses using a Green's function approach based on mode-resolved phonon properties for studying ballistic, hydrodynamic, and diffusive transport dynamics. The package provides a flexible, easy-to-use, and extensive platform for modeling phonon transport physics through Python programming.

## 1. Introduction

Modeling phonon dynamics is crucial for advanced applications including thermoelectric energy conversion[1], cooling of microprocessors[2] and power electronics[3], thermal management of batteries[4], and thermal barrier coatings[5], just to name a few. The past few decades have witnessed significant advancements in phonon modeling using *ab initio* density functional theory (DFT) and atomistic simulations[6, 7]. The developments in density functional perturbation theories in the 1990s enabled predicting phonon dispersions, anharmonic linewidths, and frequency shifts once the elements and lattice structures are known[8-10]. It was not until 2007 that Broido *et al.*[11] integrated mode-resolved specific heat, group velocities, and lifetimes from DFT calculations and reported the intrinsic thermal conductivity of silicon with relaxation time approximation (RTA), in excellent agreements with experiments free of empirical parameters. However, RTA treats all phonon-phonon interactions as resistive and neglected collective transport behaviors, resulting in underestimated thermal conductivity in materials with strong normal processes. The iterative solution of the Boltzmann transport equation (BTE) implemented in the ShengBTE package by Li and Mingo et al.[12] has been widely used to compute thermal conductivity of a wide range of materials, including dielectric solids[13], two-dimensional materials[14], and wide bandgap semiconductors[15, 16]. ShengBTE package is further extended by Feng and Ruan[17] to enable four-phonon interaction events. These advances have led to the discovery of novel semiconductors with ultrahigh thermal conductivity such as cubic BAs[17-21] and BN[22]. Recent theoretical efforts have unified the phonon gas model with the Allen-Feldman theory of disordered solids, using either Wigner transport equations[23] or Kubo's formalism of linear responses[24]. These new frameworks include both the quasiparticle transport regime and the coherent transport regime and successfully explain the glasslike temperature dependence in thermal conductivity of complex and strongly anharmonic crystals[25]. In comparison with thermal conductivity prediction, simulating temperature and heat flux responses in nanostructures and with localized heat sources often involves simplifying the multidimensional BTE to tractable problems, such as using the Debye approximation for phonon dispersions and the gray approximation for phonon lifetimes [26, 27]. Hua et al. derived the Green's function of BTE using the exact phonon dispersion and mode-resolved phonon lifetimes, but the collision matrix is simplified using RTA[28, 29]. However, RTA misses the entire hydrodynamic transport regime due to the neglected momentum conservation of normal processes. Exact solution of BTE with fully *ab initio* phonon properties and full scattering matrix for modeling hydrodynamic temperature waves have been done by Chiloyan *et al.* [30]. Inverting the full scattering matrix is, however, extremely expensive, especially with a dense mesh in the Brillouin zone. A computationally efficient way to solve BTE while capturing the ballistic, hydrodynamic, and diffusive transport regimes remains to be developed.

On the other hand, molecular dynamics (MD) based phonon modeling has grown extensively due to the following reasons. First, MD imposes no cutoffs in anharmonicity orders, and captures both the particle and wavy nature of phonons. MD can access length scales up to ten micrometers [31], and timescales up to a few milliseconds with classical computational architectures [32] and even second timescale with GPU-CPU architectures [33], showing much higher computational efficiencies than DFT. In addition, MD naturally incorporates detailed structures such as defects and boundaries. These features make MD a useful tool in predicting the thermal conductivity of materials with complex atomic structures, including amorphous materials, polymers, interfaces, and nanostructures. Due to the limited availability and accuracies in empirical potentials, MD simulations have long been used for qualitative or semi-quantitative phonon modeling. With the recent burgeon of machine learning potential (MLPs) [34, 35], accuracies in MD simulations are now approaching DFT but with much lower computational demands [36]. In addition, the development of modal analysis methods has enabled extracting spectral or mode-resolved phonon properties including frequencies, transmissions, and lifetimes, such as the spectral heat flux decomposition[37], normal mode analysis[38, 39], and spectral energy density[40]. Modal analysis of MD, however, is often a nontrivial and even challenging task. Modal analysis typically involves decomposing the MD trajectories to phonon eigenmodes, such that phonon frequencies and eigenvectors need to be used as input information. Unfortunately, lattice dynamics packages with general interatomic potentials are not available. For example, the widely used GULP package [41] supports a limited set of potential functionals, compared with the diverse potential models implemented in MD simulation packages, such as LAMMPS[42]. The lack of lattice dynamics tools for generic interatomic potential models has largely inhibited performing detailed analysis of mode-resolved phonon dynamics, as well as crosschecking MD results with BTE solutions with the same set of interatomic potential.

Here we present an open-source Python package, API Phonons, for convenient modeling of phonon dynamics. Using the powerful syntax of Python, API phonons provide functions and modules for complex phonon modeling involving different packages developed for specific purposes, including LAMMPS[42] and Quippy [43] for MD simulations, Phonopy[44] for lattice dynamics, and ShengBTE[12] for thermal conductivity and phonon lifetime calculations. API Phonons enabled complex phonon calculations, including (1) extracting harmonic and anharmonic force constants from arbitrary interatomic potentials, which can be used for solving Boltzmann transport equations; (2) predicting thermal conductivity using the linear response theory (Kubo formalism), covering both quasiparticle transport and inter-band coherent transport; and (3) modeling of ultrafast pump-probe thermal responses using a Green's function approach based on mode-resolved phonon properties for studying ballistic, hydrodynamic, and diffusive transport dynamics. API Phonons provides a flexible, easy-to-use, and extensive platform for modeling phonon transport. This set of codes along with example scripts are available at GitHub [45].

## 2. Software Overview

The workflow of API Phonons is illustrated in Figure 1. We have provided a series of Python modules for interfacing atomistic simulation programs such as LAMMPS and Quippy with force constants (FCs) calculators, including Phonopy [44], Thirdorder.py, and Fourthorder.py. The Phonopy package can be used to compute harmonic force constants (FC2), dynamical matrices, and phonon dispersions. `Thirdorder.py` and `Fourthorder.py` are Python libraries for computing third-order and fourth-order force constants (FC3 and FC4) implemented in the `Thirorder.py` and `Fourthorder.py` scripts [12, 46]. All these FCs are written in formats compatible with the widely used BTE solver packages such as ShengBTE and FourPhonon. The output files from ShengBTE or FourPhonon can also serve as inputs for the `Kappa_Kubo.py` and `BTE_GreensFunction.py` for post-processing. The `Kappa_Kubo.py` implements the quasi-harmonic Green-Kubo (QHGK) method[24] based on Kubo's linear response theory for computing both the quasiparticle contribution and coherent contributions to thermal conductivity. The coherent thermal conductivity is especially important in strongly anharmonic and complex crystals with a considerable fraction of vibration modes beyond the Ioffe-Regel transition limit [47, 48] that cannot be modeled using the quasiparticle picture. The `BTE_GreensFunction.py` module provides functions computing the thermal responses to pulsed point heat sources, which can be described by Dirac functions. By integrating Green's functions of BTE over the heating and sensing profiles, thermal responses in a variety of pump-probe experiments can be effectively calculated using the `Pump_Probe.py` module, which enables studying ultrafast thermal relaxation behavior in the ballistic, hydrodynamic, and diffusive transport regimes. In the following Section 3, details of performing lattice dynamics with arbitrary potential, computing thermal conductivity from linear response theory, and ultrafast thermal responses using Green's function formalism will be discussed comprehensively.

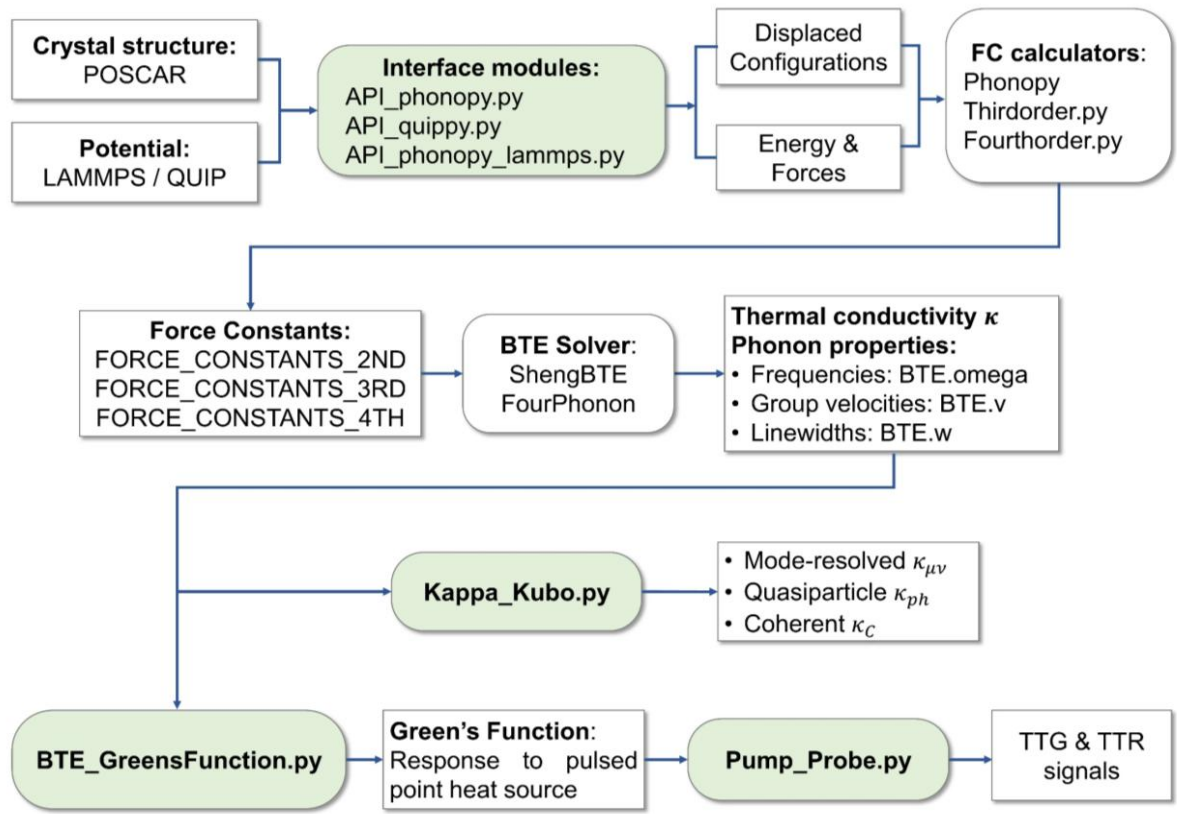

**Figure 1. The workflow of API Phonons.** The green-colored round-corner boxes are Python modules provided by API Phonons, other round-corner boxes represent phonon modeling packages. Sharp-cornered boxes represent input or output files for each step of calculations.

## 3. Methodology and Computational Details

In Section 3, we describe the methodology, and computational details, along with example scripts for performing lattice dynamics with either empirical or machine learning potential (Section 3.1), calculating thermal conductivity from linear response theory (Section 3.2), or predicting pump-probe thermal responses using the Green's function formalism of BTE (Section 3.3).

## 3.1 Lattice Dynamics with Empirical or Machine Learning Potential

Extracting FCs is crucial for the analyzing thermodynamic properties of materials and the transport of heat energy in solids. FCs are derivatives of the potential energy surface (PES) around the equilibrium configuration:

$$V = V_0 + \frac{1}{2!}\phi_{ij}^{\alpha\beta}u_i^\alpha u_j^\beta + \frac{1}{3!}\phi_{ijk}^{\alpha\beta\gamma}u_i^\alpha u_j^\beta u_k^\gamma + \frac{1}{4!}\phi_{ijkl}^{\alpha\beta\gamma\delta}u_i^\alpha u_j^\beta u_k^\gamma u_l^\delta + \cdots, \quad (1)$$

where $\phi_{ij}^{\alpha\beta}, \phi_{ijk}^{\alpha\beta\gamma}, \phi_{ijkl}^{\alpha\beta\gamma\delta}$ are harmonic, third-order, and fourth-order force constants with the subscripts indicating atoms and the superscripts representing directions ($x, y$, and $z$) in Cartesian coordinates. The $\boldsymbol{u}_i = \boldsymbol{r}_i - \boldsymbol{R}_i^0$ is the atomic displacement from the equilibrium site $\boldsymbol{R}_i^0$. Using the finite displacement method [49], FC2 is computed using the atomic forces in response to atomic displacements:

$$\phi_{ij}^{\alpha\beta} = -\frac{1}{2\Delta}\left[F_j^\beta(u_i^\alpha = \Delta) - F_j^\beta(u_i^\alpha = -\Delta)\right] \quad (2)$$

$$\phi_{ijk}^{\alpha\beta\gamma} = -\frac{1}{4\Delta^2}\Big[F_k^\gamma\left(u_i^\alpha = \Delta, u_j^\beta = \Delta\right) - F_k^\gamma\left(u_i^\alpha = \Delta, u_j^\beta = -\Delta\right) \\ - F_k^\gamma\left(u_i^\alpha = -\Delta, u_j^\beta = \Delta\right) + F_k^\gamma\left(u_i^\alpha = -\Delta, u_j^\beta = -\Delta\right)\Big] \quad (3)$$

$$\phi_{ijkl}^{\alpha\beta\gamma\delta} = -\frac{1}{8\Delta^3}\Big[F_l^\delta\left(u_i^\alpha = \Delta, u_j^\beta = \Delta, u_k^\gamma = \Delta\right) - F_l^\delta\left(u_i^\alpha = \Delta, u_j^\beta = \Delta, u_k^\gamma = -\Delta\right) \\ - F_l^\delta\left(u_i^\alpha = \Delta, u_j^\beta = -\Delta, u_k^\gamma = \Delta\right) - F_l^\delta\left(u_i^\alpha = \Delta, u_j^\beta = -\Delta, u_k^\gamma = -\Delta\right) \\ - F_l^\delta\left(u_i^\alpha = -\Delta, u_j^\beta = \Delta, u_k^\gamma = \Delta\right) - F_l^\delta\left(u_i^\alpha = -\Delta, u_j^\beta = \Delta, u_k^\gamma = -\Delta\right) \\ + F_l^\delta\left(u_i^\alpha = -\Delta, u_j^\beta = -\Delta, u_k^\gamma = \Delta\right) - F_l^\delta\left(u_i^\alpha = -\Delta, u_j^\beta = -\Delta, u_k^\gamma = -\Delta\right)\Big] \quad (4)$$

Here $\Delta$ is the magnitude of displacements. The API Phonons provide scripts for calling the LAMMPS or Quippy package as force calculators, allowing performing lattice dynamics with all potential fields supported by these atomistic simulation packages.

Here we provide an example of computing FCs of NaCl crystal using an embedded ion method (EIM) potential. First, import the phonopy and API Phonons' interfacing modules, create a Phonopy object, and generate supercells with atomic displacements through the application programming interfaces provided by Phonopy:

```
from phonopy import Phonopy
import API_phonopy as api_ph
import API_phonopy_lammps as api_pl

phonon = phonopy.load(supercell_matrix=[3, 3, 3],
```

```
                            primitive_matrix='auto',
                            unitcell_filename="POSCAR")
Scells = phonon.generate_displacements()
```

The interatomic forces can then be generated using a function calc_lmp_force_sets provided by API Phonons. The set of forces can then be parsed to the Phonopy package, to generate the force constants. Here, we show an example of calculating FC2 of NaCl using the embedded ion method (EIM) potentials:

```
cmds = ["pair_style eim","pair_coeff * * Na Cl ffield.eim Na Cl"]
forces = api_pl.calc_lmp_force_sets(cmds,Scells)
phonon.forces = forces
phonon.produce_force_constants()
phonon.symmetrize_force_constants()
fc2 = phonon.force_constants
api_ph.write_ShengBTE_FC2(fc2,filename='FORCE_CONSTANTS_2ND')
```

Here the `cmds` variable collects the LAMMPS command lines specifying the interatomic potential, which is supplied to the function `calc_lmp_force_sets` to obtain the interatomic forces necessary for extracting FC2 using Eq. (2). The calculated phonon dispersion and density of states of the NaCl using EIM potential is shown in Figure 2. Detailed scripts along with tutorial descriptions are available on GitHub.

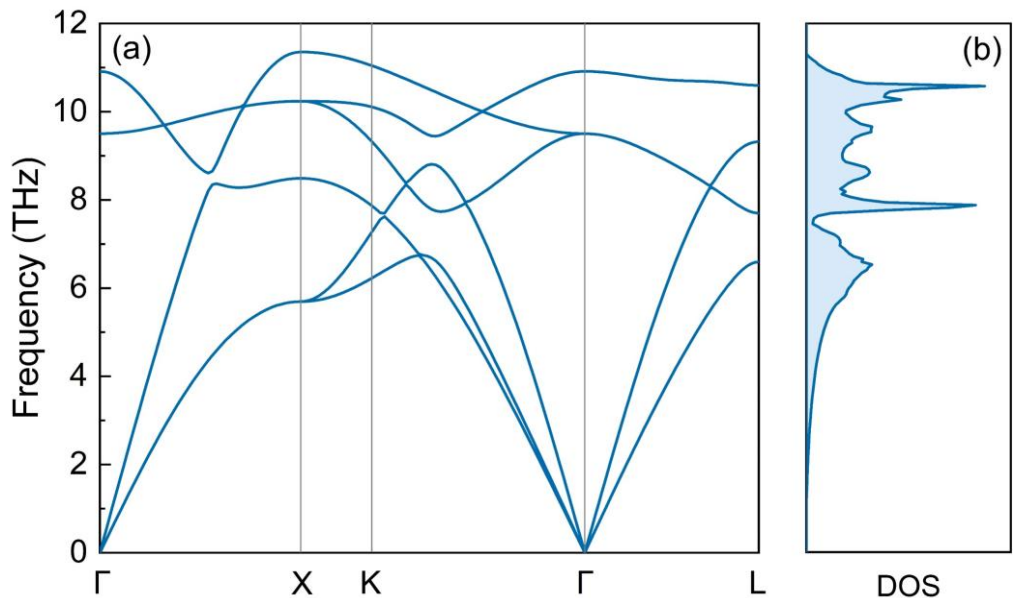

**Figure 2. Lattice dynamics of NaCl with EIM potential.** (a) Phonon dispersions and (b) density of states (DOS).

Similarly, anharmonic FC3 and FC4 can be computed by feeding atomic forces and displaced supercells to `thirdorder.py` and `fourthorer.py` packages. Here, we show an example of performing anharmonic lattice dynamics with MLP. To do this, one needs to compile LAMMPS as a shared library by enabling the MLP as a user package. For example, when computing the FCs using the machine learning Gaussian approximation potential (GAP), the forces corresponding to the displaced supercells can be obtained using:

```
cmds = [
"pair_style    hybrid/overlay quip ",
"pair_coeff    * * quip ./gp_new_DB2-4.xml \"Potential
xml_label=GAP_2018_8_16_-360_5_7_11_172\" 40"]
forces = api_pl.calc_lmp_force_sets(cmds,Scells)
```

Specifically for the GAP models, we have provided parallelized Python scripts for extracting FC3 and FC4 that can be simply running the following command line in a Linux shell:

```
python thirorder_gap_mp.py|fourthorder_gap_mp.py na nb nc
cutoff[nm|-interger] Nprocs gap_file
```

Here, `na, nb, nc` are supercell dimensions, and `Nprocs` specifies the number of processes when running these scripts, and gap_file specifies the potential file for the trained GAP. The cutoff for FC3 can be specified using a float number in nanometers or "`-n`" where n indicates the n-th nearest neighbor. The force calculations for different supercells are distributed among the processors using the multiprocessing library of python[50]. All these force constants are output in ShengBTE/FourPhonon formats, which can be directly used for thermal conductivity calculations.

In Figure 3, we show the thermal conductivity calculation of diamond using a machine learning GAP model developed by Rowe *et al* [51]. FC2, FC3, and FC4 are computed using a 6×6×6 supercell, and the cutoffs for FC3 and FC4 are seventh and the second nearest neighbor, correspondingly. Extracting FCs using API Phonons with MLPs is very efficient. For example, computing FC4 requires calculating the forces of 400 supercells with different displacements, and API Phonons can complete calculations of anharmonic FC4 within 10 minutes using eight threads. In Figure 3a, the phonon dispersion using the GAP model shows excellent agreement with inelastic neutral scattering and inelastic X-ray scattering measurements [52]. The thermal conductivity of diamond is further computed by iteratively solving the BTE using the ShengBTE and FourPhonon package, as shown in Figure 3b. When only three phonon interactions are considered, the thermal conductivity using FCs extracted from GAP is lower than DFT calculations [53], but it still shows a nice agreement with the experimental results [54]. When four-phonon interactions are further included in the modeling, thermal conductivity only shows negligible decreases below 300 K. Four-phonon interactions become increasingly important at higher temperatures, with a 15% decrease in thermal conductivity at 500 K. The relatively lower thermal conductivity calculated using FCs of GAP could be attributed to the different selection of pseudopotentials. The GAP model [51] used for extracting FCs is trained using DFT data with the optB88-vdw exchange-correlation functionals[55], while the reference DFT calculations[53] use LDA functionals[56], which typically predicts stiffer interatomic bonds and a slightly underestimated lattice constants[57].

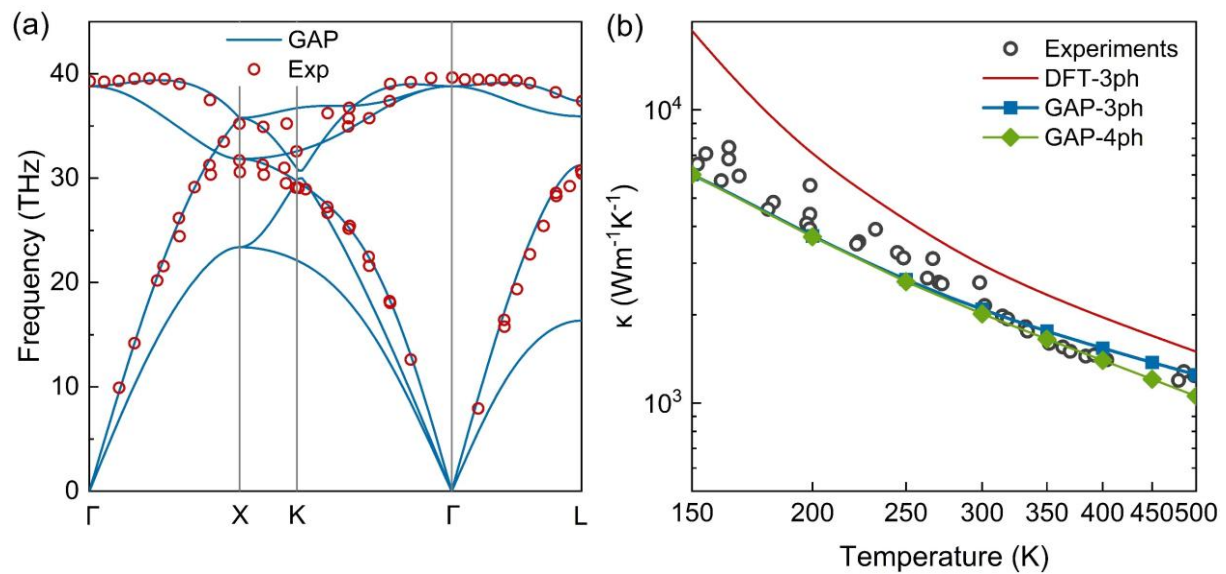

**Figure 3. Lattice dynamics of diamond with a machine learning GAP model.** (a) Phonon dispersions of the GAP model [51] compared with experiments [52]. (b) Thermal conductivity calculated with force constants extracted from the machine-learning GAP model for carbon [51], considering up to three-phonon (3ph) and four-phonon (4ph) interactions. The reference data for DFT calculations and experiments are taken from ref. [53] and ref. [54], respectively.

## 3.2 Thermal Conductivity from Linear Response Theory

In addition to providing interfaces to BTE solvers, API phonons further implement the recent QHGK method developed based on linear response theory by Isaeva *et al.* [24]. The QHGK method can predict thermal conductivities contributions from both the particle transport and the coherent interband contributions. The particle transport contributions can be simply understood with the conventional phonon gas picture. On the other hand, the coherent interband contribution is nonnegligible when the phonon linewidths are comparable with interband energy spacings. In such cases, interband tunneling can occur between phonon branches with small energy differences, as a result of phonon's wave nature[23]. The coherent interband contribution to thermal conductivity has been found significant in disordered solids[58] and strongly anharmonic crystals[59]. Alternative to QHGK, Simoncelli *et al.* also derived the formalism Wigner transport equation (WTE) [60] through quantum dynamics of density matrices that can also capture the particle-wave duality in phonon transport. While these two formalisms have different theoretical starting points, the numerical difference between QHGK and WTE has been shown as high-order and negligible[61]. In the following discussions, we outline the theoretical framework for computing thermal conductivity using the QHGK method and provide a calculation example.

Based on quantum linear response theory, the thermal conductivity is proportional to the autocorrelation function of heat fluxes[62]:

$$\kappa_{\alpha\beta} = \frac{1}{VT} \int_0^{1/k_B T} d\lambda \int_0^\infty \langle \hat{J}_\alpha(t + i\hbar\lambda)\hat{J}_\beta(0) \rangle dt \tag{5}$$

where $k_B$ and $\hbar$ are the Boltzmann constant and reduced Planck constant respectively; the subscripts $\alpha$ and $\beta$ denote directions in Cartesian coordinates, $V$, $T$ are the volume and temperature of the simulation system. $\hat{J}_\alpha(t)$ denotes the quantum heat flux operator in the Heisenberg picture, expressed as:

$$\hat{J}_\alpha(t) = \frac{i\hbar}{2} \sum_{\mu\nu} v^\alpha_{\mu\nu} \omega_\nu [\hat{a}^+_\mu(t) + \hat{a}_\mu(t)][\hat{a}^+_\nu(t) - \hat{a}_\nu(t)] \tag{6}$$

The subscripts $\mu$ and $\nu$ denote phonon mode indices, $\hat{a}^+$ and $\hat{a}$ operators are bosonic creation and annihilation operators, $\omega$ is the angular phonon frequency. The $v^\alpha_{\mu\nu}$ is the velocity operator, which can be calculated as:

$$v^\alpha_{\mu\nu} = \frac{i\delta_{q_\mu,q_\nu}}{\sqrt{\omega_\mu \omega_\nu}} \left\langle \mu \left| \frac{\partial \mathbf{D}}{\partial q_\alpha} \right| \nu \right\rangle, \tag{7}$$

where $\mathbf{D}$ is the phonon dynamical matrix, $|\nu\rangle$ and $\langle\mu|$ are the eigenvector of mode $\mu$ and the conjugated eigenvector of mode $\nu$ respectively, and $\delta_{q_\mu,q_\nu}$ is the Kronecker symbol for the wavevector of the phonon modes $\mu$ and $\nu$. By evaluating Eq. (5) using the quantum correlations $\langle \hat{a}_\mu(t)\hat{a}^+_\nu(0)\rangle = \delta_{\mu\nu}(n_\mu + 1)e^{i\omega_\mu t - \gamma_\mu t}$ and $\langle \hat{a}^+_\mu(t)\hat{a}_\nu(0)\rangle = \delta_{\mu\nu} n_\mu e^{-\gamma_\mu t + i\omega_\mu t}$, the thermal conductivity $\kappa_{\alpha\beta}$ is finally calculated as:

$$\kappa_{\alpha\beta} = \sum_{\mu\nu} \kappa^{\alpha\beta}_{\mu\nu} = \sum_{\mu\nu} C_{\mu\nu} v^\alpha_{\nu\mu} v^\beta_{\mu\nu} \tau_{\mu\nu} \tag{8}$$

where the $\kappa^{\alpha\beta}_{\mu\nu} = C_{\mu\nu} v^\alpha_{\nu\mu} v^\beta_{\mu\nu} \tau_{\mu\nu}$ is the mode-pair-resolved thermal conductivity; $C_{\mu\nu}$ and $\tau_{\mu\nu}$ are the generalized volumetric heat capacity and the resonant time between phonon modes $\mu$ and $\nu$, respectively. $C_{\mu\nu}$ and $\tau_{\mu\nu}$ are expressed as:

$$C_{\mu\nu} = \frac{\hbar \omega_\mu \omega_\nu}{VT} \frac{n_\mu - n_\nu}{\omega_\mu - \omega_\nu} \tag{9}$$

$$\tau_{\mu\nu} = \frac{\gamma_\mu + \gamma_\nu}{(\omega_\mu - \omega_\nu)^2 + (\gamma_\mu + \gamma_\nu)^2} \tag{10}$$

where $n = [\exp(\hbar\omega/k_B T) - 1]^{-1}$ is the Bose-Einstein distribution, and $\gamma_\mu$ is the phonon linewidth which can be computed using Fermi's Golden rule[63]. Eq. (8) includes both the contributions from the quasiparticle picture (diagonal elements with $\mu = \nu$) and from the resonant interband contributions (off-diagonal elements $\mu \neq \nu$). Note that the generalized Eq. (10) assumed single-mode linewidths when considering interband coherences. However, when collective phonon hydrodynamics occurs, Eq. (10) is no longer valid because one has to consider the co-relaxation between different phonon modes [61]. Hydrodynamic QHGK is not implemented yet in API Phonons. The quasiparticle contribution $\kappa_P$ and the coherent interband contribution $\kappa_C$ can be therefore calculated as the trace and the sum of the off-diagonal elements:

$$\kappa_P = \mathrm{tr}[\kappa_{\mu\nu}] \tag{11}$$

$$\kappa_C = \sum_{\mu \neq \nu} \kappa_{\mu\nu} \tag{12}$$

The QHGK method outlined above is implemented in the `Kappa_Kubo.py` module. `Kappa_Kubo.py` serves as a post-processing module by reading phonon scattering rates from ShengBTE outputs. In practice, the following Python lines can be run:

```python
import Kappa_Kubo as Kubo
from phonopy import Phonopy
T = 300
Nrepeat = [2,1,2]
mesh = [7,5,8]
FC2_file = 'FORCE_CONSTANTS_2ND'
phonon = phonopy.load(supercell_matrix=Nrepeat,
                     unitcell_filename='POSCAR',
                     force_constants_filename= FC2_file)
phonon.run_mesh(mesh,is_gamma_center=True)
scatt_rate_ph = Kubo.read_ShengBTE_scattRate('BTE.w',phonon)
Kappa_Kubo,Kappa_P,Kxx_mp,Kyy_mp,Kzz_mp,freqs =
Kubo.calc_QHGK_ShengBTE_at_T(phonon,mesh,scatt_rate_ph,T)
```

`Kappa_Kubo` and `Kappa_P` are respectively the thermal conductivity evaluated using the Kubo formalism and the quasiparticle BTE theory. `Kxx_mp,Kyy_mp,Kzz_mp` corresponds to $\kappa_{\mu\nu}^{xx}$, $\kappa_{\mu\nu}^{yy}$, and $\kappa_{\mu\nu}^{zz}$ for all mode pairs $(\mu, \nu)$ along the three Cartesian coordinates. Note that the `mesh` variable is the Brillouzone q-mesh for computing phonon lifetimes, which should be set consistent with ShengBTE calculations. Jupyter notebook files with interactive Python cells for computing thermal conductivity using the `Kappa_Kubo.py` are available on GitHub.

Figure 4 summarizes a case study of quasiparticle and coherent transport of phonons in a strongly anharmonic material CsPbBr$_3$ (*Pnma* group) using the `Kappa_Kubo.py` module of API Phonons. CsPbBr$_3$ has a perovskite structure with a complex unit cell[64], whose thermal conductivity cannot be well described by the BTE under the quasiparticle picture[23]. Harmonic phonon properties of CsPbBr$_3$ are obtained from the Materials Data Repository database[65], and the anharmonic FC3 is computed using the finite-displacement method with a cutoff up to the fourth nearest neighbor. Phonon lifetimes are then computed using ShengBTE on a Brillouin zone mesh of 7×5×8, consistent with Simoncelli *et al.* [23]. As shown in Figure 4a, the phonon dispersions of CsPbBr$_3$ are featured by lots of nearly flat bands closely bundled together. When the frequency separations are close to or smaller than phonon linewidths, the vibrational modes can no longer be regarded as particles, and the coherent interband contribution becomes nonnegligible. As shown in Figure 4b, the quasiparticle contribution $\kappa_P$ obtained from BTE significantly underestimates the thermal conductivity. $\kappa_P$ features a $T^{-1}$ temperature dependence, consistent with the predictions by the quasiparticle picture. On the other hand, the coherent interband conductivity $\kappa_C$ increases at high temperatures, because larger phonon linewidths result in a larger resonant lifetime $\tau_{\mu\nu}$ and thereby the off-diagonal conductivities $\kappa_{\mu\nu}$, as shown Figure 4c-d. At 50 K, the coherent interband $\kappa_C$ contributes to only 17% of $\kappa_{tot}$; the coherent $\kappa_C$ contributes to more than 50% of $\kappa_{tot}$ at 300 K. After the coherent interband contributions $\kappa_C$ are considered, the predicted thermal conductivity $\kappa_{tot} = \kappa_P + \kappa_C$ shows an excellent agreement with the experimental reports [66] from cryogenic temperatures to room temperature. The predicted $\kappa_{tot}$ using QHGK is ~15% higher than experimental reports at the low temperature 50 K. At such low temperature, phonon transport and thermal conductivity become very sensitive to defects and boundaries, which is not included in QHGK modeling. In addition, it remains to be studied that how pseudopotentials for computing FCs can affect the relative contributions from $\kappa_P$ and $\kappa_C$. The major computational cost of QHGK modeling implemented here lies in calculating phonon lifetimes, and API Phonons serve as a post-processing script. In this example of phonon coherent transport in CsPbBr$_3$, running the Python scripts only takes 10 seconds without any parallelization.

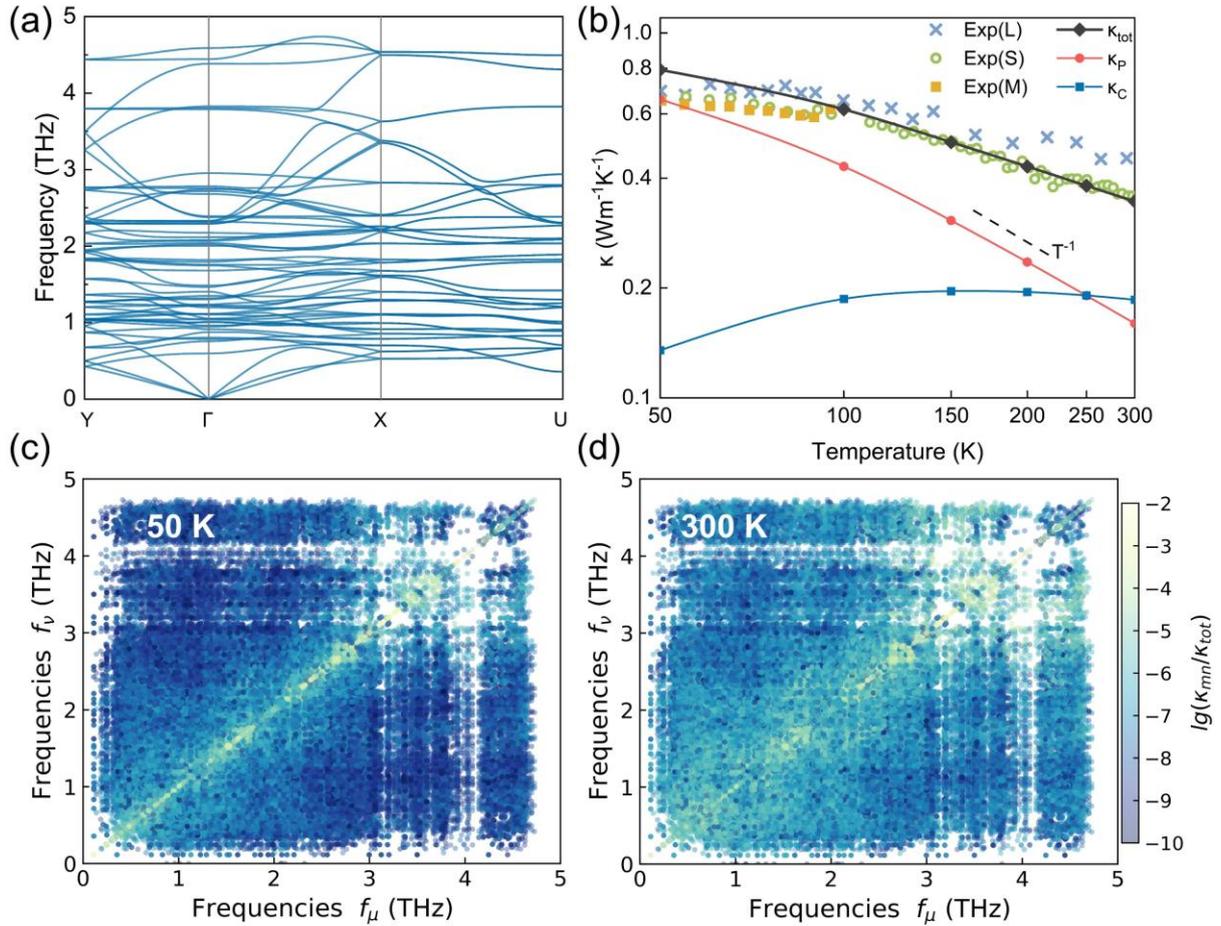

**Figure 4. Thermal conductivity of CsPbBr₃ calculated by the QHGK method.** (a) Phonon dispersion of CsPbBr₃. (b) Temperature dependence in the total thermal conductivity $\kappa_{tot}$, the quasiparticle contributions $\kappa_P$ and the interband tunneling contribution $\kappa_C$ along the [100] direction. Exp(L), Exp(S), and Exp(M) refer to nanowire samples with different cross-sections [66], and their agreement with each other indicates negligible boundary scattering compared with phonon-phonon scatterings. The mode-resolved thermal conductivity $\kappa_{\mu\nu}$ at (c) 50 K and (d) 300 K.

## 3.3 Ultrafast Pump-Probe Thermal Responses using Green's Function of BTE

Thermal phonon transport significantly deviates from the diffusive Fourier regime when the characteristic length of the system is comparable to the phonon mean free paths or when the characteristic heating frequency is comparable with the phonon scattering rate. Temperature and heat flux distributions in these nondiffusive transport regimes are governed by the BTE. Solving BTE with Green's function approach provides a simple, closed-form method for calculating thermal responses in infinite or semi-infinite domains under arbitrary shapes of heat inputs. API Phonons implements the Green's function solutions of BTE under the Callaway scattering approximation (CSA) [67], which separately treats relaxation times of the momentum conserving and reversing processes. Such CSA treatment is a more physical representation of detailed phonon-phonon scatterings than RTA, which allows modeling transport dynamics in the hydrodynamic transport regimes where phonon-phonon interaction is dominated by N-scattering events. In contrast, the RTA treats all phonon-phonon interactions as resistive, and misses the entire hydrodynamic transport regime as a result. CSA also avoids the demanding computational costs of rigorously solving BTE by inverting the full scattering matrix. Here we outline the computational approach for obtaining Green's functions, detailed derivation can be found in ref. [68].

To obtain Green's function, we consider the mode-dependent BTE under a unit pulsed point heat source described by Dirac functions $\delta(\bm{r})\delta(t)$:

$$\frac{\partial n_\mu}{\partial t} + \bm{v}_\mu \cdot \nabla n_\mu = \frac{n_\mu^0(T_R) - n_\mu}{\tau_\mu^R} + \frac{n_\mu^d(T_N) - n_\mu}{\tau_\mu^N} + \frac{p_\mu}{\hbar \omega_\mu}\delta(\bm{r})\delta(t), \tag{13}$$

where $n_\mu$ is the nonequilibrium phonon distribution, $\bm{v}_\mu = \nabla_{\bm{q}}\omega_\mu$ is the group velocity ($\omega_\mu$ is the angular frequency of phonon mode $\mu$); $p_\mu = C_\mu/C$ is the fraction of energy distributed to mode $\mu$, with $C_\mu$ and $C$ the modal and total specific heat. $\tau_\mu^N$ and $\tau_\mu^R$ are respectively the relaxation time for momentum-conserving N-scattering processes and momentum-reversing R-scattering events. $n_\mu^0$ and $n_\mu^d$ are correspondingly the local equilibrium and the drifted equilibrium distributions:

$$n_\mu^0(T_R) = \frac{1}{\exp\left(\frac{\hbar\omega_\mu}{k_B T_R(\bm{r})}\right) - 1} \tag{14}$$

$$n_\mu^d(T_N, \bm{u}) = \frac{1}{\exp\left[\frac{\hbar(\omega_\mu - \bm{q}_\mu \cdot \bm{u})}{k_B T_N(\bm{r})}\right] - 1} \tag{15}$$

Here $\bm{u}$ is the collective drift velocity, $T_N$ and $T_R$ are pseudo-temperatures for the local equilibrium of N-scattering and R-scattering events respectively. In the near-equilibrium transport regime, the energy deviations can be analytically solved in the Fourier-transform domain:

$$g_\mu(\Omega, \boldsymbol{\xi}) = \chi_\mu \left( C_\mu [\rho_\mu \Delta T_R(\Omega, \boldsymbol{\xi}) + \eta_\mu \Delta T_N(\Omega, \boldsymbol{\xi})] + C_\mu T_0 \eta_\mu \frac{\boldsymbol{q} \cdot \boldsymbol{u}}{\omega_\mu} + p_\mu \tau_\mu \right), \quad (16)$$

where $g_\mu = \hbar \omega_\mu [n_\mu - n_\mu^0(T_0)]$ is the energy deviation from equilibrium with a baseline temperature $T_0$. $\rho_\mu = \tau_\mu/\tau_\mu^R$ and $\eta_\mu = \tau_\mu/\tau_\mu^N$ are the relative strengths of R- and N-scattering normalized by the total scattering rate, respectively. The variables $(\Omega, \boldsymbol{\xi})$ are the temporal and spatial frequencies from Fourier transforms: $(t, \boldsymbol{r}) \to (\Omega, \boldsymbol{\xi})$. $\chi_\mu$ is the phonon susceptibility defined as:

$$\chi_\mu = \frac{1}{1 + i\Omega\tau_\mu + i\boldsymbol{F}_\mu \cdot \boldsymbol{\xi}}, \quad (17)$$

where $\boldsymbol{F}_\mu$ is the mean free displacement of phonon mode $\mu$. The unknown pseudo-temperatures and the unknown drift velocity $\boldsymbol{u}$ can be solved with the energy and momentum conservation requirements:

$$\sum_\mu \frac{1}{\tau_\mu^R} (C_\mu \Delta T_R - g_\mu) = 0 \quad (18)$$

$$\sum_\mu \frac{1}{\tau_\mu^N} (C_\mu \Delta T_N - g_\mu) = 0 \quad (19)$$

$$\sum_\mu \frac{\hbar \boldsymbol{q}}{\tau_\mu^N} [n_\mu - n_\mu^d(T_N)] = 0 \quad (20)$$

Eqs (18-20) can be written in the following linear form:

$$\boldsymbol{A}\boldsymbol{X} = \boldsymbol{b} \quad (21)$$

The linear matrix $\boldsymbol{A}$ takes the form:

$$\boldsymbol{A} = \begin{bmatrix} \sum_\mu \frac{C_\mu}{\tau_\mu^R}\left(1 - \frac{\tau_\mu}{\tau_\mu^R}\chi_\mu\right) & -\sum_\mu C_\mu \frac{\tau_\mu}{\tau_\mu^R \tau_\mu^N}\chi_\mu & -\sum_\mu \frac{C_\mu T_0 \boldsymbol{q}^{\mathrm{T}}}{\omega_\mu}\frac{\tau_\mu}{\tau_\mu^R \tau_\mu^N}\chi_\mu \\ -\sum_\mu C_\mu \frac{\tau_\mu}{\tau_\mu^R \tau_\mu^N}\chi_\mu & \sum_\mu \frac{C_\mu}{\tau_\mu^N}\left(1 - \frac{\tau_\mu}{\tau_\mu^N}\chi_\mu\right) & \sum_\mu \frac{C_\mu T_0 \boldsymbol{q}^{\mathrm{T}}}{\omega_\mu \tau_\mu^N}\left(1 - \frac{\tau_\mu}{\tau_\mu^N}\chi_\mu\right) \\ -\sum_\mu \frac{C_\mu \boldsymbol{q}}{\omega_\mu}\frac{\tau_\mu}{\tau_\mu^R \tau_\mu^N}\chi_\mu & \sum_\mu \frac{C_\mu \boldsymbol{q}}{\omega_\mu \tau_\mu^N}\left(1 - \frac{\tau_\mu}{\tau_\mu^N}\chi_\mu\right) & \sum_\mu \frac{C_\mu T_0 \boldsymbol{qq}}{\omega_\mu^2 \tau_\mu^N}\left(1 - \frac{\tau_\mu}{\tau_\mu^N}\chi_\mu\right) \end{bmatrix}, \quad (22)$$

The vector $\boldsymbol{X}$ is the responses in pseudo-temperatures and drift velocities, while the vector $\boldsymbol{b}$ collects energy generation rates for R- and N-processes, and the momentum generation rate for the N processes:

$$\boldsymbol{X} = \begin{bmatrix} \Delta T_R \\ \Delta T_N \\ \boldsymbol{u} \end{bmatrix}, \qquad \boldsymbol{b} = \begin{bmatrix} \sum_\mu \frac{\tau_\mu}{\tau_\mu^R} p_\mu \\ \sum_\mu \frac{\tau_\mu}{\tau_\mu^N} \chi_\mu p_\mu \\ \sum_\mu \frac{\tau_\mu}{\tau_\mu^N} \chi_\mu p_\mu \boldsymbol{q}/\omega_\mu \end{bmatrix}. \quad (23)$$

The unknown vector can be simply solved as $\boldsymbol{X} = \boldsymbol{A}^{-1}\boldsymbol{b}$. Finally, the local temperature response can be obtained from the total energy conservation:

$$\sum_\mu \frac{1}{\tau_\mu}(C_\mu \Delta T - g_\mu) = 0, \quad (24)$$

Together with Eq. (18-19), Green's function of the local energy deviation is derived as:

$$\mathcal{G}_{\Delta T} = \sum_\mu C_\mu \left(\frac{X_R}{\tau_\mu^R} + \frac{X_N}{\tau_\mu^N}\right) \Big/ \sum_\mu C_\mu \tau_\mu^{-1}, \quad (25)$$

where $X_R$ and $X_N$ denote the first two components of $\boldsymbol{X}$. In the R-scattering limit ($\tau_\mu^N \to \infty$, $\tau_\mu \approx \tau_\mu^R$), the Green's function can be simply calculated as:

$$\mathcal{G}_{\Delta T}^{RTA} = \frac{\sum_\mu \chi_\mu p_\mu}{\sum_\mu C_\mu \tau_\mu^{-1}(1 - \chi_\mu)}, \quad (26)$$

which is identical to the analytical results obtained by Hua et al [28].

The Green's function $\mathcal{G}_{\Delta T}$ embeds the temperature response to the unit heat source, with which temperature rise to any heating profile $P(\Omega, \boldsymbol{\xi})$ can be simply calculated as $\Delta T(\Omega, \boldsymbol{\xi}) =$ $\mathcal{G}_{\Delta T}(\Omega, \xi) P(\Omega, \xi)$. For pump-probe experiments, the measured temperature rise is weighted by a sensing profile $S(\Omega, \xi)$, and the frequency-domain thermal signal is calculated as:

$$\Delta \bar{T}(\Omega) = \int \mathcal{G}_{\Delta T}(\Omega, \xi) P(\Omega, \xi) S(\Omega, \xi) d\xi \qquad (27)$$

Time-domain responses can then be simply calculated using inverse Fourier transformation.

In practice, the `BTE_GreensFunction.py` and the `Pump_Probe.py` modules need to be imported first as shown below:

```python
import BTE_GreensFunction as BTEGF
import Pump_Probe as PuPr
```

These two modules provide functions for calculating the Green's function and for integrating over the pump and probe profiles. For example, when computing transient thermal grating (TTG) signals, one needs to generate meshgrids in the temporal and spatial domains and then compute (or load) the Green's function using the function `get_BTEGFs`. The Python code looks like:

```
XIx, OMEGAH = BTEGF.Generate_TTGMeshGrid(L,FreqH_MHz)

Meshgrid,GdT,Gu,GdT_RTA = BTEGF.get_BTEGFs(T0,load_GFs,is_isotope,
 (XIx,OMEGAH),qpoints_full,phonon,rots_qpoints,freqs,cqs,vqs,Fqs,
 tau_qs,tauN_qs,tauR_qs)
```

The variables `L` and `FreqH_MHz` for generating meshgrids represent the grating period and the vector of heating frequencies, respectively. When calling `get_BTEGFs`, one needs to specify the equilibrium temperature (`T0`), the full list of points for the q-mesh in the brillouin zone (`qpoints_full`), the phonon frequencies (`freqs`), mode-resolved specific heats (`cqs`), group velocities (`vqs`), mean-free-displacements (`Fqs`), relaxation times (`tau_qs`), and the relaxation times for N and U processes (`tauN_qs,tauR_qs`). The material structure for obtaining the phonon properties is grouped in the `phonon`, which is a phonopy object. The bool-type variable `load_GFs` specifies whether one needs to load the existing data (True) or recompute the Green's function (False). The `is_isotope` specifies whether the isotope scattering is included in `tauR_qs`, which helps select correct data when loading the existing data. In the output variables, `GdT`, `Gu`, and `GdT_RTA` are correspondingly the Green's functions of temperature rise $\mathcal{G}_{\Delta T}$, the collective drift velocity $\mathcal{G}_u$, and the temperature rises at the RTA limit.

After Green's functions have been obtained, functions in `Pump_Probe.py` can be called to calculate different pump-probe geometries. Figure 5a shows the schematic of the pump-probe TTG experiment as an example. Two pulsed pump beams are cross-focused at the sample surface, generating an interference heating pattern. The surface heat source therefore has a sinusoidal profile: $P(t,r) \propto \cos(2\pi x/L)\,\delta(t)$, with $L$ the grating period that diffracts the probe beam. Such spatially periodic heating would modulate the optical properties of the sample and create a transient optical grating that will diffract the probe beam. In the heterodyne detection of TTG signals, an extra reference beam is introduced, whose reflection is overlapped with the diffracted probe beam to amplify the signal [69]. In the Fourier domain, the heating profile corresponds to two Dirac delta functions centered at $\xi_x = \pm 2\pi/L$, hence the detected TTG signal is calculated as:

$$\Delta \bar{T}(t) = \frac{1}{4\pi} \int_{-\infty}^{\infty} [\mathcal{G}_{\Delta T}(\Omega, \xi = 2\pi/L) + \mathcal{G}_{\Delta T}(\Omega, \xi = -2\pi/L)] e^{i\Omega t} d\Omega \quad (28)$$

API Phonons provide functions to predict pump-probe signals. The ultrafast TTG responses can be calculated using the following Python code:

```
OmegaH = 2*np.pi*FreqH_MHz/1e6 # heating frequencies in THz
t,Tt,wH,Tw = PuPr.calc_TTGSig(XIx,OmegaH,GdT,Tmax,np.max(OmegaH))
```

In the above code block, `Tmax` specifies the upper bound in time when performing inverse Fourier transform. The output variables `t`, `Tt` are correspondingly the time and the transient temperature response; and `wH`, `Tw` denotes the frequency-domain representations of the time-domain signals.

Figure 5b shows the TTG signal predicted using Green's functional formalism, using phonon properties of graphite obtained using the ShengBTE package with *ab initio* FCs and a Brillouin zone mesh of 24×24×10. The Green's function approaching with CSA successfully captures the transient hydrodynamic lattice cooling phenomena, *i.e.* the negative dip in the normalized temperature below zero [70, 71]. In quasiballistic or Fourier transport regimes, temperature decays monotonically after a heating pulse. In the hydrodynamic regime, phonons transport collectively and the propagation of temperature disturbance becomes wavelike. As a result, the temperature can even decrease below the initial temperature when the oscillation phase is shifted by $\pi$, which is the signature of the second sound. Figure 5c shows the TTG response in the frequency domain, which is the Fourier transform of the time-domain responses. The peak in the frequency domain hallmarks the formation of the second sound. At elevated temperatures(Figure 5d), the temperature decay becomes monotonically decreasing, and the BTE solution approaches the Fourier limit. The resonant peak of the second sound also vanishes in the frequency-domain responses at the elevated temperature of 250 K (Figure 5e).

This analytical Green's function formalism implemented in API Phonons is more efficient than numerical solutions of BTE. For example, the discrete ordinate method (DOM) typically has the complexity of $\mathcal{O}(N_\mu N_r N_\theta N_t)$ [72], where $N_\mu$ is the total number of phonon states; $N_r$, $N_\theta$, and $N_t$ are the number of descrte spatial coordinates, number of discrete propagation angles of phonons, and the number of timesteps, respectively. Solving the BTE using the Monte Carlo method typically has a similar complexity to DOM, but it requires a few independent runs to reduce the stochastic fluctuations through ensemble average. In Green's function formalism, the full BTE is reduced to a linear system $\boldsymbol{AX} = \boldsymbol{b}$, with $\boldsymbol{A}$ being only a small 6×6 matrix at each Fourier-domain coordinate $(\Omega, \boldsymbol{\xi})$. While evaluating the matrix $\boldsymbol{A}$ has the complexity of $\mathcal{O}(N_\mu N_r N_t)$, the complexity of the key step obtaining the Green's function is reduced $\mathcal{O}(N_r N_t)$, independent of the number of phonon modes. In comparison to our analytical Green's function using CSA, computing Green's function by inverting the full scattering matrix ($N_\mu \times N_\mu$) becomes very demanding, with the computational complexity of $\mathcal{O}(N_r N_t N_\mu^3)$. As a result, computing the hydrodynamic TTG signals takes less than 1 minute, even without any parallelization. Another advantage is that, once Green's function is calculated, one can save the data of Green's function and use it to compute phonon transport from different geometries of heat sources simply using Eq. (27) without solving the BTE again.

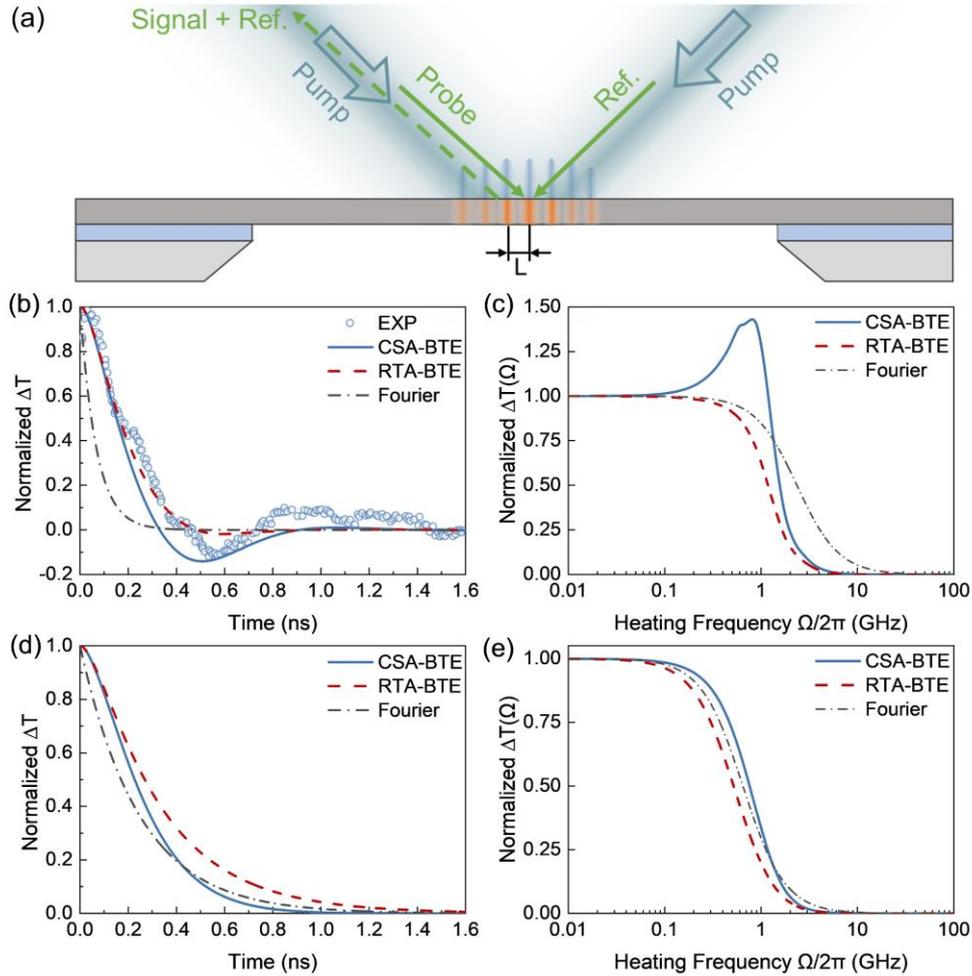

**Figure 5. Predicted TTG signals using Green's function formalism.** (a) Schematic for the generation of TTG and the detection of diffracted signal. Two pump beams are cross-focused on the sample surface to create a transient grating, and the diffracted probe beam is detected. An extra reference beam (ref.) is introduced whose reflection is overlapped with the diffracted probe to amplify the TTG signal. (b) Comparisons in TTG signals predicted using CSA, RTA, and Fourier theory and (c) The frequency-domain thermal responses normalized by steady-state temperature rise, with a grating period of 4 μm at 100 K. (d) TTG signals and the (e) frequency-domain responses at 250 K with the same grating period.

## 4. Summary and Future Developments

We have presented API Phonons, a Python software for modeling thermal phonon transport. The software streamlines complex calculation tasks involving different packages developed for specific purposes, including LAMMPS and Quippy for atomistic simulations, Phonopy for lattice dynamics, and ShengBTE/FourPhonon for computing thermal conductivity and anharmonic phonon properties. The package enabled extracting harmonic and anharmonic force constants from arbitrary interatomic potentials supported by LAMMPS and Quippy, predicting both the quasiparticle and coherent contributions to thermal transport, and modeling ultrafast pump-probe thermal responses using mode-resolved phonon properties. API Phonons provide an extensive and easy-to-use platform for studying the ballistic, hydrodynamic, and diffusive transport dynamics of phonons.

Finally, we comment on the limitations and future developments of API Phonons. The current Python package is not parallelized using OpenMP/MPI. For more complicated pump-probe experiments such as time-domain thermoreflectance (TDTR), parallelization becomes necessary because predicting TDTR signals would require computing and summing $10^4 \sim 10^5$ frequency components[73]. Future developments on API Phonons will parallelize the modules computing Green's function of BTE for computationally demanding modeling of pump-probe experiments. API Phonons has not implemented interfacial transport modeling yet. Modules computing transmission spectra across hetero-interfaces and the interfacial thermal transport in modeling pump-probe measurements will be implemented in the future.

**Acknowledgment**

X.Q. acknowledges the support from the National Key R & D Project (Grant No. 2022YFA1203100). T.H.L acknowledges support from the National Natural Science Foundation of China (NSFC, Grant No. 52076089).

## Data Availability

The software, interactive Python script examples, and raw data have been posted on GitHub:

https://github.com/xinqian-mit/API_Phonons.

## References

1. X.-L. Shi, J. Zhou and Z.-G. Chen, Advanced Thermoelectric Design: From Materials and Structures to Devices, Chem. Rev. **120** (15), 7399-7515 (2020).
2. E. Pop, S. Sinha and K. E. Goodson, Heat Generation and Transport in Nanometer-Scale Transistors, Proc. IEEE **94** (8), 1587-1601 (2006).
3. S. V. Garimella, A. S. Fleischer, J. Y. Murthy, A. Keshavarzi, R. Prasher, C. Patel, S. H. Bhavnani, R. Venkatasubramanian, R. Mahajan, Y. Joshi, B. Sammakia, B. A. Myers, L. Chorosinski, M. Baelmans, P. Sathyamurthy and P. E. Raad, Thermal Challenges in Next-Generation Electronic Systems, IEEE Transactions on Components and Packaging Technologies **31** (4), 801-815 (2008).
4. M. T. Agne, T. Böger, T. Bernges and W. G. Zeier, Importance of Thermal Transport for the Design of Solid-State Battery Materials, PRX Energy **1**, 031002 (2022).
5. D. R. Clarke and S. R. Phillpot, Thermal Barrier Coating Materials, Mater. Today **8** (6), 22-29 (2005).
6. A. J. H. McGaughey, A. Jain and H.-Y. Kim, Phonon Properties and Thermal Conductivity from First Principles, Lattice Dynamics, and the Boltzmann Transport Equation, J. Appl. Phys. **125** (1), 011101 (2019).
7. A. J. H. McGaughey, University of Michigan, 2004.
8. P. Giannozzi, S. de Gironcoli, P. Pavone and S. Baroni, Ab Initio Calculation of Phonon Dispersions in Semiconductors, Phys Rev B Condens Matter **43** (9), 7231-7242 (1991).
9. J. M. Ziman, *Electrons and Phonons: The Theory of Transport Phenomena in Solids*. (Oxford University Press, 2001).
10. A. Debernardi, S. Baroni and E. Molinari, Anharmonic Phonon Lifetimes in Semiconductors from Density-Functional Perturbation Theory, Phys. Rev. Lett. **75** (9), 1819-1822 (1995).
11. D. A. Broido, M. Malorny, G. Birner, N. Mingo and D. A. Stewart, Intrinsic Lattice Thermal Conductivity of Semiconductors from First Principles, Appl. Phys. Lett. **91** (23), 231922 (2007).
12. W. Li, J. Carrete, N. A. Katcho and N. Mingo, Shengbte: A Solver of the Boltzmann Transport Equation for Phonons, Comput. Phys. Commun. **185** (6), 1747-1758 (2014).
13. S. Stackhouse, L. Stixrude and B. B. Karki, Thermal Conductivity of Periclase (Mgo) from First Principles, Phys. Rev. Lett. **104** (20), 208501 (2010).
14. X. Gu, Y. Wei, X. Yin, B. Li and R. Yang, Phononic Thermal Properties of Two-Dimensional Materials, arXiv:1705.06156 (2017).
15. N. H. Protik, A. Katre, L. Lindsay, J. u. Carrete, N. Mingo and D. Broido, Phonon Thermal Transport in 2h, 4h, and 6h Silicon Carbide from First Principles, Materials Today Physics **1**, 31-38 (2017).
16. L. Lindsay, D. A. Broido and T. L. Reinecke, Thermal Conductivity and Large Isotope Effect in Gan from First Principles, Phys. Rev. Lett. **109** (9), 095901 (2012).
17. T. Feng, L. Lindsay and X. Ruan, Four-Phonon Scattering Significantly Reduces Intrinsic Thermal Conductivity of Solids, Phys. Rev. B **96** (16), 161201(R) (2017).

18. F. Tian, B. Song, X. Chen, N. K. Ravichandran, Y. Lv, K. Chen, S. Sullivan, J. Kim, Y. Zhou, T.-H. Liu, M. Goni, Z. Ding, J. Sun, G. A. G. U. Gamage, H. Sun, H. Ziyaee, S. Huyan, L. Deng, J. Zhou, A. J. Schmidt, S. Chen, C.-W. Chu, P. Y. Huang, D. Broido, L. Shi, G. Chen and Z. Ren, Unusual High Thermal Conductivity in Boron Arsenide Bulk Crystals, Science **361**, 582-585 (2018).
19. J. S. Kang, M. Li, H. Wu, H. Nguyen and Y. Hu, Experimental Observation of High Thermal Conductivity in Boron Arsenide, Science **361**, 575-578 (2018).
20. S. Li, Q. Zheng, Y. Lv, X. Liu, X. Wang, P. Y. Huang, D. G. Cahill and B. Lv, High Thermal Conductivity in Cubic Boron Arsenide Crystals, Science **361**, 579-581 (2018).
21. L. Lindsay, D. A. Broido and T. L. Reinecke, First-Principles Determination of Ultrahigh Thermal Conductivity of Boron Arsenide: A Competitor for Diamond?, Phys. Rev. Lett. **111** (2), 025901 (2013).
22. K. Chen, B. Song, N. K. Ravichandran, Q. Zheng, X. Chen, H. Lee, H. Sun, S. Li, G. A. G. U. Gamagan, F. Tian, Z. Ding, Q. Song, A. Rai, H. Wu, P. Koirala, A. J. Schmidt, K. Watanabe, B. Lv, Z. Ren, L. Shi, D. G. Cahill, T. Taniguchi, D. Broido and G. Chen, Ultrahigh Thermal Conductivity in Isotope-Enriched Cubic Boron Nitride, Science **367**, 555-559 (2020).
23. M. Simoncelli, N. Marzari and F. Mauri, Unified Theory of Thermal Transport in Crystals and Glasses, Nature Physics **15** (8), 809-813 (2019).
24. L. Isaeva, G. Barbalinardo, D. Donadio and S. Baroni, Modeling Heat Transport in Crystals and Glasses from a Unified Lattice-Dynamical Approach, Nature communications **10** (1), 3853 (2019).
25. Y. Xia, V. Ozolins and C. Wolverton, Microscopic Mechanisms of Glasslike Lattice Thermal Transport in Cubic $Cu_{12}sb_4s_{13}$ Tetrahedrites, Phys. Rev. Lett. **125** (8), 085901 (2020).
26. C. Zhang, S. Huberman and L. Wu, On the Emergence of Heat Waves in the Transient Thermal Grating Geometry, J. Appl. Phys. **132** (8), 085103 (2022).
27. C. Zhang, S. Chen and Z. Guo, Heat Vortices of Ballistic and Hydrodynamic Phonon Transport in Two-Dimensional Materials, Int. J. Heat Mass Transfer **176**, 121282 (2021).
28. C. Hua and A. J. Minnich, Analytical Green's Function of the Multidimensional Frequency-Dependent Phonon Boltzmann Equation, Phys. Rev. B **90** (21), 214306 (2014).
29. J. Xu, Y. Hu and H. Bao, Quantitative Analysis of Nonequilibrium Phonon Transport near a Nanoscale Hotspot, Physical Review Applied **19** (1), 014007 (2023).
30. V. Chiloyan, S. Huberman, Z. Ding, J. Mendoza, A. A. Maznev, K. A. Nelson and G. Chen, Green's Functions of the Boltzmann Transport Equation with the Full Scattering Matrix for Phonon Nanoscale Transport Beyond the Relaxation-Time Approximation, Phys. Rev. B **104**, 245424 (2021).
31. G. Barbalinardo, Z. Chen, H. Dong, Z. Fan and D. Donadio, Ultrahigh Convergent Thermal Conductivity of Carbon Nanotubes from Comprehensive Atomistic Modeling, Phys. Rev. Lett. **27**, 025902 (2021).
32. John L Klepeis, K. Lindorff-Larsen, R. O. Dror and D. E. Shaw, Long-Timescale Molecular Dynamics Simulations of Protein Structure and Function, Current Opinion in Structural Biology **19** (2), 120-127 (2009).
33. J. C. Phillips, D. J. Hardy, J. D. C. Maia, J. E. Stone, J. V. Ribeiro, R. C. Bernardi, R. Buch, G. Fiorin, J. Hénin, W. Jiang, R. McGreevy, M. C. R. Melo, B. K. Radak, R. D. Skeel, A. Singharoy, Y. Wang, B. Roux, A. Aksimentiev, Z. Luthey-Schulten, L. V. Kalé, K. Schulten, C. Chipot and E. Tajkhorshid, Scalable Molecular Dynamics on Cpu and Gpu Architectures with Namd, The Journal of chemical physics **153** (4), 044130 (2020).
34. T. Mueller, A. Hernandez and C. Wang, Machine Learning for Interatomic Potential Models, The Journal of chemical physics **152** (5), 050902 (2020).

35. V. Botu, R. Batra, J. Chapman and R. Ramprasad, Machine Learning Force Fields: Construction, Validation, and Outlook, J. Phys. Chem. C **121** (1), 511-522 (2017).
36. D. Dragoni, T. D. Daff, G. Csányi and N. Marzari, Achieving Dft Accuracy with a Machine-Learning Interatomic Potential: Thermomechanics and Defects in Bcc Ferromagnetic Iron, Physical Review Materials **2** (1), 013808 (2018).
37. Y. Zhou and M. Hu, Full Quantification of Frequency-Dependent Interfacial Thermal Conductance Contributed by Two- and Three-Phonon Scattering Processes from Nonequilibrium Molecular Dynamics Simulations, Phys. Rev. B **95**, 115313 (2017).
38. A. Henry and G. Chen, Spectral Phonon Transport Properties of Silicon Based on Molecular Dynamics Simulations and Lattice Dynamics, Journal of Computational and Theoretical Nanoscience **5**, 1-12 (2008).
39. A. J. H. McGaughey and J. M. Larkin, Predicting Phonon Properties from Equilibrium Molecular Dynamics Simulations, Annual Review of Heat Transfer **17**, 49-87 (2014).
40. J. A. Thomas, J. E. Turney, R. M. Iutzi, C. H. Amon and A. J. H. McGaughey, Predicting Phonon Dispersion Relations and Lifetimes from the Spectral Energy Density, Phys. Rev. B **81** (8), 081411(R) (2010).
41. J. D. Gale, Gulp: A Computer Program for the Symmetry-Adapted Simulation of Solids, J. Chem. Soc., Faraday Trans **93**, 629-637 (1997).
42. S. Plimpton, Fast Parallel Algorithms for Short-Range Molecular Dynamics, Journal of Computational Physics **117**, 1-19 (1995).
43. G. Csanyi, S. Winfield, J. Kermode, M. C. Payne, A. Comisso, A. D. Vita and N. Bernstein, Expressive Programming for Computational Physics in Fortran 950+, Newsletter of the Computational Physcis Group, 1-24 (2007).
44. A. Togo and I. Tanaka, First Principles Phonon Calculations in Materials Science, Scripta Mater. **108**, 1-5 (2015).
45. Api Phonons, https://github.com/xinqian-mit/API_Phonons.
46. Z. Han, X. Yang, W. Li, T. Feng and X. Ruan, Fourphonon: An Extension Module to Shengbte for Computing Four-Phonon Scattering Rates and Thermal Conductivity, Comput. Phys. Commun. **270**, 108179 (2022).
47. S. N. Taraskin and S. R. Elliott, Ioffe-Regel Crossover for Plane-Wave Vibrational Excitations in Vitreous Silica, Phys. Rev. B **61**, 12031 (2000).
48. Y. Luo, X. Yang, T. Feng, J. Wang and X. Ruan, Vibrational Hierarchy Leads to Dual-Phonon Transport in Low Thermal Conductivity Crystals, Nature communications **11** (1), 2554 (2020).
49. G. Kresse, J. Furthmuller and J. Hafner, Ab Initio Force Constant Approach to Phonon Dispersion Relations of Diamond and Graphite, Europhys. Lett. **32** (9), 729-734 (1995).
50. https://docs.python.org/3/library/multiprocessing.html.
51. P. Rowe, V. L. Deringer, P. Gasparotto, G. Csányi and A. Michaelides, An Accurate and Transferable Machine Learning Potential for Carbon, The Journal of chemical physics **153**, 034702 (2020).
52. M. Schwoerer-Böhning, A. T. Macrander and D. A. Arms, Phonon Dispersion of Diamond Measured by Inelastic X-Ray Scattering, Phys. Rev. Lett. **80** (25), 5572 (1998).
53. A. Ward, D. A. Broido, D. A. Stewart and G. Deinzer, Ab Initiotheory of the Lattice Thermal Conductivity in Diamond, Phys. Rev. B **80** (12), 125203 (2009).
54. J. R. Olson, R. O. Pohl, J. W. Vandersande, A. Zoltan, T. R. Anthony and W. F. Banholzer, Thermal Conductivity of Diamond between 170 and 1200 K and the Isotope Effect, Phys. Rev. B **47** (22), 14850-14856 (1993).
55. J. Klimeš, D. R. Bowler and A. Michaelides, Van Der Waals Density Functionals Applied to Solids, Phys. Rev. B **83** (19), 195131 (2011).

56. J. P. Perdew and A. Zunger, Self-Interaction Correction to Density-Functional Approximations for Many-Electron Systems, Phys. Rev. B **23** (10), 5048-5079 (1981).
57. P. Haas, F. Tran and P. Blaha, Calculation of the Lattice Constant of Solids with Semilocal Functionals, Phys. Rev. B **79**, 085104 (2009).
58. P. B. Allen and J. L. Feldman, Thermal Conductivity of Disordered Harmonic Solids, Phys. Rev. B **48** (17), 12581-12588 (1993).
59. T. Zhu and E. Ertekin, Mixed Phononic and Non-Phononic Transport in Hybrid Lead Halide Perovskites: Glass-Crystal Duality, Dynamical Disorder, and Anharmonicity, Energy & Environmental Science **12** (1), 216-229 (2019).
60. M. Simoncelli, N. Marzari and F. Mauri, Wigner Formulation of Thermal Transport in Solids, Physical Review X **12** (4), 041011 (2022).
61. A. Fiorentino and S. Baroni, From Green-Kubo to the Full Boltzmann Kinetic Approach to Heat Transport in Crystals and Glasses, Phys. Rev. B **107** (5) (2023).
62. R. Kubo, M. Yokota and S. Nakjima, Statistical-Mechanical Theory of Irreversible Processes. Ii. Response to Thermal Disturbance, J. Phys. Soc. Jpn. **12** (11), 1203-1211 (1957).
63. A. Debernardi, Phonon Linewidth in Iii-V Semiconductors from Density Functional Perturbation Theory, Phys. Rev. B **57**, 12847 (1998).
64. C. C. Stoumpos, C. D. Malliakas, J. A. Peters, Z. Liu, M. Sebastian, J. Im, T. C. Chasapis, A. C. Wibowo, D. Y. Chung, A. J. Freeman, B. W. Wessels and M. G. Kanatzidis, Crystal Growth of the Perovskite Semiconductor Cspbbr3: A New Material for High-Energy Radiation Detection, Crystal Growth & Design **13** (7), 2722-2727 (2013).
65. https://mdr.nims.go.jp/concern/datasets/wh246x832?locale=en.
66. Y. Wang, R. Lin, P. Zhu, Q. Zheng, Q. Wang, D. Li and J. Zhu, Cation Dynamics Governed Thermal Properties of Lead Halide Perovskite Nanowires, Nano Lett. **18** (5), 2772-2779 (2018).
67. J. Callaway, Model for Lattice Thermal Conductivity at Low Temperatures, Phys. Rev. **113** (4), 1046-1051 (1959).
68. X. Qian, C. Zhang, T.-H. Liu and R. Yang, Analytical Green's Function of Multidimensional Boltzmann Transport Equation for Modeling Hydrodynamic Second Sound, arXiv: 2410.20146 (2024).
69. J. A. Johnson, A. A. Maznev, M. T. Bulsara, E. A. Fitzgerald, T. C. Harman, S. Calawa, C. J. Vineis, G. Turner and K. A. Nelson, Phase-Controlled, Heterodyne Laser-Induced Transient Grating Measurements of Thermal Transport Properties in Opaque Material, J. Appl. Phys. **111** (2), 023503 (2012).
70. S. Huberman, R. A. Duncan, K. Chen, B. Song, V. Chiloyan, Z. Ding, A. A. Maznev, G. Chen and K. A. Nelson, Observation of Second Sound in Graphite at Temperaturers above 100 K, Science **364** (6438), 375-379 (2019).
71. Z. Ding, K. Chen, B. Song, J. Shin, A. A. Maznev, K. A. Nelson and G. Chen, Observation of Second Sound in Graphite over 200 K, Nature communications **13**, 285 (2022).
72. Y. Hu, Y. Shen and H. Bao, Optimized Phonon Band Discretization Scheme for Efficiently Solving the Nongray Boltzmann Transport Equation, J. Heat Transfer **144**, 072501 (2022).
73. D. G. Cahill, Analysis of Heat Flow in Layered Structures for Time-Domain Thermoreflectance, Rev. Sci. Instrum. **75** (12), 5119-5122 (2004).